\documentstyle[12pt]{article}
\begin{document}
\baselineskip .3in
\begin{titlepage}
\begin{center}{\large{\bf On the Production of Jets and Hadrons }}
\vskip .2in
S.Mukherjee(nee Banerjee ) $^{\dag}$  \\
\vskip .1in School of Mathematical Sciences,
\vskip .01in Dublin Institute of Technology, \\
Dublin,Ireland.\\

\end{center}

\vskip .3in

{\bf Abstract}

The ratio of the semi­vertical cone angles of the quark and gluon
jets calculated in the statistical model is found to be in exact
agreement with that derived from QCD.In hadron production also,the
model's prediction of the power law type of behaviour for the
inclusive distribution is strikingly close to the corresponding
experimental findings;these spectacular results are significant
for the model to be realistic. \vskip .1in PACS:
13.87.Ce,13.87.Fh,13.85.Ni,13.66.Bc

\vskip .3in

\noindent $^{\dag}$ E-mail:smukherjee@dit.ie

\end{titlepage}

\newpage
 \vskip .1in
{\bf 1.Introduction}

It is well­known that colour forces organise quarks and gluons
into hadrons through the fragmentic process.Becasue of the
non-perturbative processes dominating it,exact theoretical
prediction is still lacking and only predictions of model­type
calculations are available.A characteristic property of many of
the most  interesting events in hadron collider data is the
appearance of jets which are dominated by short­distance effects
so that the perturbative Quantum Chromodynamics(pQCD) may be
safely applied to study jet processes.As the hard quark scattering
is manifest through the jet production;it becomes essential to
study its  inclusive production and to analyse the scattering at
the largest possible transverse momenta($p_{t}$)scale possible.In
high energy pp interactions,the measurements of $\pi^{0}$ meson
spectra in a high $p_{t}$ ($ p_{t}  > 2GeV/c$) domain show a
strong deviation from the exponential shape and Gazdzicki et.
al[1] have found that there is the power law in hadron production
and the mean multiplicity and transverse mass spectra obey such a
law.Moreover,the estimate of the parameters in the power law are
found to be the same when these are adjusted against corresponding
experimental data for the production of different mesons at fixed
collision energy.The results of our investigations on the jet
observables,cross sections for particle production etc. have been
compared with the relevant experimental findings and the
predictions of pQCD,which help us to analyse the current status of
the Statistical Model. \vskip .01in The Statistical Model of the
hadron as a quarkonium system was proposed more than a couple of
decades ago[2] and has been widely used to study the properties of
hadrons.Since,the model has also undergone modifications[3]along
with its various ramifications and applications[3,4]. In the
original version of the model[2] to be referred hereafter as the
model (A),no allowance has been made for the exchange effect and
we come across the fractal dimension $D = \frac{9}{2}$ of the
hadron.However,for comparitively moderate values of the momenta of
the constituents of a hadron so that the exchange energy effect
may play an appreciable role [2],we arrive at D=8.5 in the
modified version of the model(B).In the current investigation ,the
ratio $\frac{9}{4}$ expressed in terms of the angular radii of the
quark and gluon jets,has been derived only from the geometric
properties of the hadron in the framework of the Statistical
Model(A).Moreover,the analysis of the cross sections for
production of jets and hadrons reveals the power law falling of as
$p_{t}^{-D}$ where $p_{t}$ is the transverse momentum of the
secondary and this is found to be in accord with the experimental
findings for moderate as well as for asymptotically large values
of $p_{t}$ .\vskip .01in
It may be relevant to recall that the
boundary of a jet reveals a hierarchy of indentations with a
complex local structure and it behaves like a fractal [5] and the
concept of fractal trees was imbibed to model the jets that arise
when particles collide head on at very high energy.Hence it would
be interesting to study if such a behaviour also exists in
hadronic jets as the differences between the quark and gluon jets
are also known to originate from QCD and it is curious to know if
such a pattern is also manifest through the geometrical
properties.Since the quarks and gluons are viewed as carriers of
colour charges,the asymptotic prediction of the ratio  as that of
the average parton multiplicates in jets can thus be related to
the charged hadron multiplicities[6].However,without using the
assumption of the parton model,Sterman and Weinberg[7] have
investigated the properties of hadronic jets in $e^{+}e^{-}$
annihilation in QCD and have asserted that their properties follow
as consequences of pQCD and thus Einhorn and Weeks [8] have used
QCD to investigate the properties of jets produced by a gluon
source and have derived the relation between the jet opening
angles .Shiuya and Tye[9] have also calculated the jet angular
radius  of a gluon jet in the framework of the pQCD and have
estimated the ratio of the quark jet angular radius ($\delta_{q}$)
to that of the gluon jet ($\delta_{g}$).Defining the fraction f of
all jet like events which have a fraction ($1 - \epsilon^{'} $) of
the total energy E inside some pair of opposite cones of
half­ang.$\delta $ we have,
\begin{equation}
f(gluon) = 1 - \frac{\alpha(E)}{\pi}[4C_{2}(G)ln
2\epsilon^{'}+\frac{11}{3} C_{2}(G) - \frac{4}{3}
N_{f}T(R)]ln\delta
\end{equation}

The same formula for the quark case [7] is

\begin{equation}
f(quark) = 1 - \frac{\alpha(E)}{\pi}[4ln2 \epsilon^{'}+
3]C_{2}(R)ln \delta
\end{equation}

where the colour factor $C_{2}(R) = \frac{4}{3} $,$C_{2}(G) =
3$and $T(R) = \frac{1}{2}$ for a quark triplet and all other
symbols have their usual meanings.For the same values of
f,E,$\epsilon^{'}$ ,the ratio of the angular radii of the quark
and gluon jets is independent of the coupling constant $\alpha(E)$
and in the asymptotic limit of $p^{2}$ ,the first­order prediction
for the jet angular sizes yields the ratio[8,9] as
\begin{equation}
  \frac{ln \delta_{q}}{ln \delta_{g}}\rightarrow
  \frac{C_{2}(G)}{C_{2}(R)}=\frac{9}{4}
\end{equation}
where $ \delta_{q}$ and $\delta_{g}$ are in radians.On the other
hand,in our model for the quark jet,we have a cone of
semi­vertical angle $\delta_{q}$ where the volume of the right
circular cone dimensions is $\frac{\pi h^{3}}{3}tan^{2}\delta_{q}$
where h is the height of the cone.Approximating $tan\delta_{q}$ by
$\delta_{q}$ for $\delta_{q}$ very small, we get
$\delta_{q}^{2}=\frac{3V}{\pi h^{3}}$ and hence in the context of
the model (A),we have in the D dimensional space of the hadron
with $\beta$ a constant factor, $\delta^{2}_{q}=\frac{\beta
V}{h^{D}}$ and we get

\begin{equation}
\frac{ln \delta_{q}}{ln \delta_{g}}= \frac{ln \delta^{2}_{q}}{ln
\delta^{2}_{g}}=\frac{ln(\beta V)/(h^{D})}{2 ln \delta_{g}}=
\frac{1}{2}[\frac{ln (\beta V/h^{D})}{ln \delta_{g}}]
\end{equation}
As in the box­counting method,we consider the quark jet to be
covered with N balls of radius r and $\frac{r}{h}= \epsilon$,
$\frac{V}{h^{D}}\sim \epsilon^{D}$ ,the unit of measurement where
$\epsilon $ is a dimensionless very small quantity ($ \epsilon \ll
1$) which is the size of the covering balls normalised by the
linear size of the strucutre.Hence we have $N(\epsilon)\sim
\epsilon^{-D}$,the number of D dimensional balls of radius
$\epsilon h$ needed to cover the fractal dimension D defined as $
\lim_{\epsilon\rightarrow  0} [\frac{ln
N(\epsilon)}{ln\epsilon^{-1}}]$. From (4) we have therefore,

\begin{equation}
\frac{ln \delta_{q}}{ln \delta_{g}}= \frac{1}{2}\frac{ln (\beta
V/h^{D})}{ln(r/h)}\rightarrow \frac{1}{2}\lim_{\epsilon\rightarrow
0}\frac{ln \epsilon^{D} }{ln \epsilon}=\frac{1}{2}
\lim_{\epsilon\rightarrow 0}\frac{ln N(\epsilon)}{ln
\epsilon^{-1}} =\frac{D}{2}=\frac{9}{4}
\end{equation}

Hence we have
\begin{equation}
\delta_{g}= (\delta_{g})^{4/9}
\end{equation}
This remarkable result was also obtained by Einhorn et al.[8] and
Shizuya et al[9] from QCD.Thus the aforesaid relation(6) which
indicates that the gluon jet spreads more than the quark jet has
been derived in the current investigation purely from
(fractal)geometrical properties of the hadron suggested by the
model. \vskip .01in As the power law behaviour can often be
attributed o the consequences of the underlying self­similarity,we
analyse its predictions of the decrease of the cross­sections of
pp, $p\overline{p}$ collisions and $e^{+}e^{-}$ annihilation in
$p_{t}$ in the context of the model.The analysis of these
experimental data for the production of particles is likely to
provide the information required to make important and stringent
tests of its internal consistency.Thus,recently we [10] have
derived and analysed the power law scaling in the light of the
model and have found that it works well with $\alpha =8.5$ as an
input corresponding to the model (B) for the transverse mass
spectra ($m_{t}$ ) of pions ($m_{t} > 1Gev/c^{2}$ )and for
particles from eta to upsilon.Our result are found to be in
excellent agreement with the corresponding experimental findings
with D = 8.5 for $p_{t}> 2 Gev/c$. On the other hand,Gazdzicki et
al[1] have indicated that the properly normalised multiplicities
and transverse mass ($m_{t}$) spectra of neutral mesons obey the
scaling law $cm_{t}^{-\alpha}$ where
$m_{t}=\sqrt{m^{2}+p_{t}^{2}}$ and c and $\alpha $  are similar
for all mesons produced at the same collision energy.Introducing
the hadronic mass spectrum )such that $\rho(m_{t} )$ such that
$\rho(m_{t} )dm_{t}$ is the number of different hadron states in
the transverse mass interval $(m_{t} , dm_{t})$ ,we have for
\begin{equation}
  \rho(m_{t})\sim p_{t}^{-\alpha}
\end{equation}
for $p_{t}\gg m$.Our assertion is that the cross­section for the
production of hadrons is proportional to the hadronic mass
spectrum with $\alpha= 8.5$ and 4.5 for moderate values and asymp­
totic large values of $p_{t}$ respectively.We would like to
assert,as in Hagedon[11],that almost all strong interactions may
be simulated through $\rho(m_{t})$ as the highly excited hadronic
matter or fireball contains undeterminate number of all species of
hadrons and as such the scaling property displayed in(57) is also
expected for jets observed in hadron collisions,jet fragmentation
etc.On the other hand,the inclusive production of pions with large
$p_{t}$ in high energy pp collisions has been extensively studied
and the invariant cross sections are also observed to obey a
scaling law of the form
\begin{equation}
  E\frac{d\sigma}{d^{3}p}\propto y(p_{t}= p_{t}^{-
  \alpha}I(x_{t},\theta)
\end{equation}
where E and p are the energy and momentum of the secondary,
$\theta$ is the production angle, $p_{t}=psin \theta$ is the
transverse momentum,the scaling variable
$x_{t}=\frac{2p_{t}}{\sqrt{s}}$ and y is a scaling function of
$p_{t}$ .In the case of pure hadronic collisions,the inclusive
cross sections of the production of secondaries having limited
$p_{t}$ with respect to the collision axis possess the Feynman
scaling and leads to the asymptotic behaviour with high $p_{t}$ in
$y(p_{t})$ as $p_{t}^{-4}$ in contradiction with available
experimental data.However,to match the quark counting role,the
cross section must have the asymptotic $p_{t}^{-8}$ (pion)if the
production of high $p_{t}$ particles is conditioned by hard
scattering of a quark from a composite system of two quarks.Field
and Feynman [12] have found that all experiments on production at
$90^{0}$ scales as in (8) with $\alpha$ lying between 7.2 and
8.6.Feynman et al.[13] have also reexamined(8) for the large
$p_{t}$ production of particles and jets in QCD approach for $pp
\rightarrow \pi +X$ at $\theta_{cm}=90^{0}$ for  some moderate
values values of $p_{t}$ and $x_{t}$ and have found that
$\alpha=8$ and the right hand side of (8) becomes a universal
function of $x_{t}$ and $p_{t}$ only and the falling of the
aforesaid experimental results according to (8) is found to be
close to our prediction(B) as $\alpha=8.5$.However,for $p_{t} =
10Gev/c$ at $x_{t} = 0.2$,it is close to $p_{t}^{-5}$ and this is
again in reasonable accord with our
prediction(A)i.e.$\alpha=4.5$.Clark et al [14]have reported
measurements of inclusive $\pi^{0}$ production from high energy pp
collisions at very large $p_{t}$ at Fermilab and at CERN ISR at
the centre of mass energies 58 and 63 Gev and $\theta\simeq
90^{0}$ and have found that the data are in accord with (8)and
that lies in the range of 7 to 9 for $x_{t}$ between 0.2 and
0.45.Experimental findings on charged pions from Fermi lab by
Antreasyan et al[15] with 200,30 and 400 GeV pp collisions for
$p_{t}$ ranging from 0.77 to 6.91 GeV/c also suggest that  lies
$\alpha $ between 8.2 and 8.5.Again,Ellis et al[16] have pointed
out that $ \sqrt{s}= 30.8GeV$ at ISR,and $\alpha \simeq 9$ and the
study by McCubbin [17] on hadron production for large $p_{t}$ in
deep inelastic scattering of hadron hadron process also reveals
that when the invariant single particle cross section is plotted
against large $p_{t}$ for over the energy range from 5 to to 63
GeV, $\alpha$ turns out to be equal to 8.3 and $\alpha$ becomes 8
when the energy ranges from 20 to 60 Gev.The steeper falling o of
the cross section for inclusive distribution up to 10 GeV/c at ISR
energies behaves as $p_{t}^{-8}$ which compares favourably with
the model's prediction of $p_{t}^{-8.5}$  .On the otherhand,at
much higher energies ,the $p_{t}^{-4.5}$ type of asymptotic
dependence of $y(p_{t})$ is suggested by the model (A) and
obviously it compares reasonably well with the corresponding QCD
of $p_{t}^{-4}$ and the experimental findings.Further in the
trigger bias effect,regarded as a very important consequence of
scaling in the jet fragmentation,the associated distribution on
the trigger side is also found to scale in the particle
distribution $\frac{d\sigma}{dp_{t}}$ as proportional to
$p_{t}^{-9}$ compared to the model's prediction (B) of
$p_{t}^{-8.5}$ when $\frac{d\sigma}{dp_{t}}$ is assumed to be
proportional to $\rho (m _{t})$.\vskip .01in In studying single
particle cross sections of the process $pp \rightarrow \pi + X$ at
$\theta_{cm}=90^{0}$,Feynman et al[13] have observed its
$p_{t}^{-8}$ behaviour at $p_{t} \leq 6 Gev/c$ at $ x_{t}= 0.2$
and at $p_{t} \leq 10 Gev/c $ at $ x_{t}= 0.5$ along with its
asymptotic approach like $p_{t}^{-4}$ .We may assert,therefore,in
view of the above data that $y(p_{t})$ has the cross­ over point $
p_{t}^{c}$ of the $p_{t}$ variable where $p_{t}^{c}= 6GEV/c $ or
$10 GeV/c$ for the two cases respectively.As different exponents
(8.5 and 4.5)are needed on either side of $p_{t}^{c}$ for both the
cases as in a self­affine curve,the invariance of the measure is
recovered by the affine type of transformation as in self­affine
fractals. The ratio of the semi­vertical angles of the quark and
gluon jets in the asymptotic limit coincides exactly with that
predicted from QCD;moreover ,the power law decrease of the
inclusive cross sections as $p_{t}^{-8.5}$ in
pp,$p\overline{p}$,collisions and in $e^{+}e^{-}$ annihilation is
consistent with the corresponding experimental
findings.Further,for asymptotically large values of $p_{t}$ ,the
aforesaid power law becomes $p_{t}^{-4.5}$ which is not only in
accord with the QCD prediction but it also agrees with the
corresponding experimental data which fall approximately as
$p_{t}^{-4}$ .The model credited with the successes of these
striking predictions and agreements with the experimental findings
vouches for its realistic character for the description of nature.

\vskip .1in  {\bf Acknowledgements}

The author is thankful to Prof.S.N.Banerjee,Department of
Physics,Jadavpur University,Kolkata,India for encouragement.

\newpage
\vskip .2in {\bf References}

\noindent[1] M.Gazdzicki and M.I. Gorenstein,Phys.Lett.B
517,250(2001). \noindent [2] S.N.Banerjee et. al., Had.
J.4,2003(1981);(E)5,2157(1982); 6,
440,956(1983);7,186,795(1984);J.Phys.G.8,L61(1982);
 Ann.Phys.
(N.Y.)150,150(1983); Nuov.Cim.A102,1733(1989)

\noindent [3]S.N. Banerjee et.al.,Had. J.
11,243(1988);12,179(1989);13,75(1990).

\noindent [4] S.N.Banerjee et.al.Int.
J.Mod.
Phys.A10,201(2001);
Int.J.Mod.Phys.A17,4939(2002).

\noindent [5]B.Mandelbrot,The Fractal Geometry(W.H. Freeman and
Company,New York,U.S.A;1983)98.

\noindent [6]S.J.Brodsky and J.F.
Gunion,
Phys.Rev.Lett.
37,402(1976);
Skluth,Rep.Prog.Phys.
65,1771(2006).

\noindent [7] G.Sterman and S.Weinberg,Phys.
Rev.Lett.39,1436(1977).

\noindent [8]M.B.Eiinhorn and B.G.Weeks,Nucl.Phys.B146,455(1978).

\noindent [9]K.Shizuya and S.H.H. Tye,Phys.Rev.lett 41,787(1978).

\noindent [10] S.N.Banerjee and
S.Banerjee,Phys.Lett.B644,45(2007).

\noindent [11] R.Hagedorn,CERN 71-72,38(1971).

\noindent [12] R.D.Field and R.P. Feynman,Phys.Rev.D15,2590(1977).

\noindent [13]R.P.Feynman et al.,Phys. Rev.D18,3320(1978).

 \noindent [14] A.G. Clark et al,Phys.Lett.74B,267(1978)

\noindent[15]D.Antreasyan et al.,Phys.Rev.Lett.38,112(1977).

\noindent[16]S.D.Ellis et al.,Nucl.Phys.B108,93(1976).

\noindent[17]N.A.McCubbin,Phys.Scr.25,143(1982).

\noindent[18]G.Giacomelli and M.Jacob,Phys.Rep.55,82(1979).

\noindent[19] M.D.Sharpiro and J.L.
Siegrist,Ann.Rev.Nucl.Part.Sci,41,97(1991).

\end{document}